\documentclass{Interspeech}
\usepackage{multirow}
\usepackage{makecell}
\usepackage{subcaption}

\interspeechcameraready

\title{FD-Bench: A Full-Duplex Benchmarking Pipeline Designed for Full Duplex Spoken Dialogue Systems}

\author[affiliation={1,2}]{Yizhou}{Peng}
\author[affiliation={2}]{Yi-Wen}{Chao}
\author[affiliation={2}]{Dianwen}{Ng}
\author[affiliation={3}]{Yukun}{Ma}
\author[affiliation={3}]{Chongjia}{Ni}
\author[affiliation={3}]{Bin}{Ma}
\author[affiliation={2}]{Eng Siong}{Chng}

\affiliation{Alibaba-NTU Global e-Sustainability CorpLab}{Nanyang Technological University}{Singapore}
\affiliation{College of Computing and Data Science}{Nanyang Technological University}{Singapore}
\affiliation{Alibaba}{Alibaba Inc.}{Singapore}

\email{peng.yizhou@ntu.edu.sg}
\keywords{Large Language Models, Evaluation, Full-Duplex Spoken Dialogue Systems}

\usepackage{comment}

\begin{document}

\maketitle

\begin{abstract}
Full-duplex spoken dialogue systems (FDSDS) enable more natural human-machine interactions by allowing real-time user interruptions and backchanneling, compared to traditional SDS that rely on turn-taking. However, existing benchmarks lack metrics for FD scenes, e.g., evaluating model performance during user interruptions. In this paper, we present a comprehensive FD benchmarking pipeline utilizing LLMs, TTS, and ASR to address this gap. It assesses FDSDS’s ability to handle user interruptions, manage delays, and maintain robustness in challenging scenarios with diverse novel metrics. We applied our benchmark to three open-source FDSDS (Moshi, Freeze-omni, and VITA-1.5) using over 40 hours of generated speech, with 293 simulated conversations and 1,200 interruptions. The results show that all models continue to face challenges, such as failing to respond to user interruptions, under frequent disruptions and noisy conditions. Demonstrations, data, and code will be released.\footnote{https://github.com/pengyizhou/FD-Bench}
    
\end{abstract}

\begin{figure*}[htb]
    \vspace{-0.9em}
    \centering
    \includegraphics[width=1.00\linewidth]{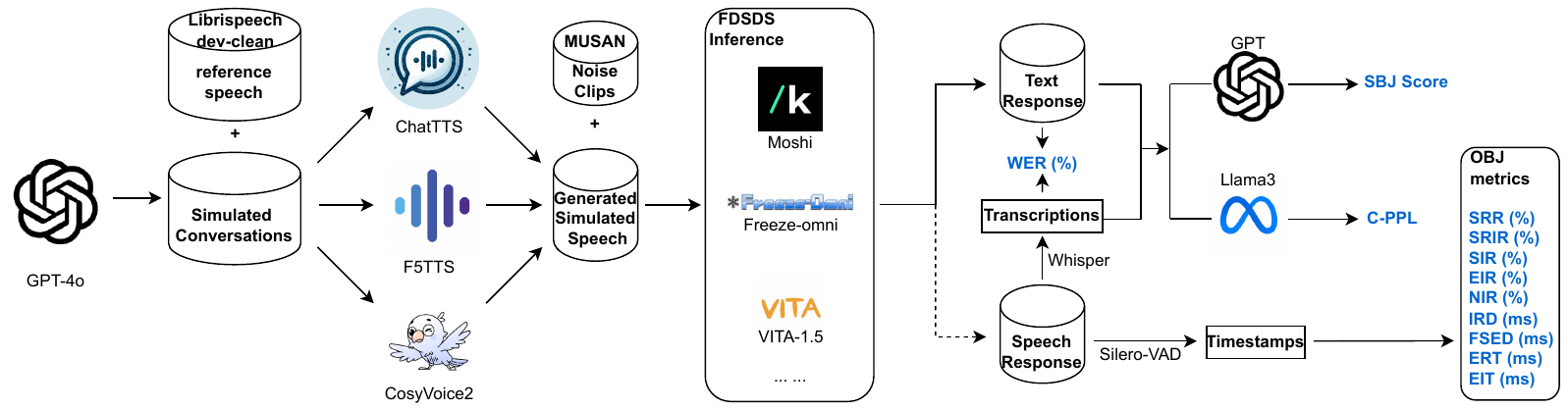}
    \vspace{-1.5em}
    \caption{
    Pipeline for Benchmarking FDSDS: The framework integrates simulated conversations generated by GPT-4o and speech synthesis tools to produce input data. Noise samples and reference speech are used for diverse speakers and environments. The pipeline processes these inputs through duplex systems, incorporating Whisper for transcriptions and Silero-VAD for obtaining timestamps. Subjective scoring involves GPT inference, while Conditioned-PPL is from Llama3. Objective metrics are computed using timestamps. 
    }
    \label{fig:bm-pipeline}
    \vspace{-1.5em}
\end{figure*}

\section{Introduction}
Spoken dialogue systems (SDS) enable natural human-machine interactions through spoken language. Traditionally, SDS were built using cascaded architectures—such as the early Google Duplex~\cite{Google-Duplex}—which relied on sequentially processing inputs through modular components (ASR, Speech Interruption Handler, NLU, and TTS), a design that, while groundbreaking, often suffered from latency issues and error accumulation. Recently, end-to-end (E2E) SDS like AudioGPT~\cite{audioGPT}, SpeechGPT~\cite{speechgpt}, Mini-Omni~\cite{mini-omni}, and GLM4~\cite{glm4-voice} have emerged, leveraging large language models (LLMs) such as pretrained Llama3~\cite{llama3}, and Qwen~\cite{qwen} with simplified architectures to reduce latency and improve accuracy significantly. Additionally, since the speed of speech generation also constrains the performance of SDS, an improved diffusion method~\cite{zhang2024speaking} focuses on achieving faster and more responsive speech synthesis. However, despite these advances, most current SDS remain turn-taking-based, which limits their ability to replicate the fluidity of human conversation. To overcome this limitation, full-duplex SDS (FDSDS) — exemplified by Moshi~\cite{Moshi}, Freeze-Omni~\cite{Freeze-omni}, OmniFlatten~\cite{OmniFlatten}, and Minmo~\cite{Minmo}—have been introduced, enabling simultaneous speaking and listening, effectively handling real-time conversational dynamics such as interruptions, affirmations, backchanneling, and topic shifts~\cite{Ma2024LanguageMC, traum-evaluating-2015-ACL}.

These developments have spurred the creation of several benchmarks that standardize system evaluation. Researchers actively assess these systems’ understanding, reasoning, and speech generation capabilities, as well as their accuracy in solving various problems~\cite{HongYee-SDS, SLAM-Omni}. Moreover, studies investigate how diverse speech styles, such as varying accents, ages, speeds, and volumes, enhance overall system performance~\cite{Benchmark2}.
However, these metrics are predominantly assessing the model performance turn-by-turn instead of a full-duplex manner. This means the users wait until the model finishes speaking before responding, which doesn’t mirror real-world interactions where people engage with one another in real time. 




In this work, we propose a comprehensive benchmarking pipeline for FDSDS that addresses the previously unmet need for full-duplex capability evaluation. It automates dialogue generation, speech corpus simulation, and multi-criteria assessment using LLMs, TTS, and ASR technologies. This end-to-end pipeline is scalable to larger datasets, making it applicable to any FDSDS without requiring manual evaluation of the model output.

This study identifies key challenges and unresolved issues by applying our benchmarking pipeline to a diverse set of state-of-the-art FDSDS, including Moshi, Freeze-omni, and VITA-1.5~\cite{VITA}. Our experiments highlight key aspects such as robustness to interruptions, latency management, and natural response delays, offering valuable insights into the current limitations of these systems. These findings underscore the need for further research on more natural, human-like spoken dialogue systems. 

\section{Methodology}
\label{sec: med}
We propose a benchmarking pipeline to evaluate the full-duplex capabilities of SDS, addressing a gap in current evaluation approaches by incorporating a thorough assessment of interruption handling. This pipeline defines full-duplex performance across two key dimensions: interruption handling and response quality. The first dimension assesses the system's ability to handle interruptions effectively, focusing on its capacity to detect user-initiated disruptions, manage ongoing processes (such as halting and resuming operations), and control delays during these events. The second dimension evaluates the system’s response quality following an interruption, ensuring that generated speech remains stable and that the conversational content stays coherent and aligned with the intended dialogue. All tests are conducted within a server-client architecture to emulate real-world usage scenarios and ensure reliable, consistent performance metrics.

\subsection{Benchmarking Pipeline}

Figure~\ref{fig:bm-pipeline} shows our overall pipeline of benchmarking designed for FDSDS. We first use GPT-4o to generate simulated conversational text, up to 5 conversation rounds, between a user and an AI agent. Then, we utilize State-of-the-Art speech synthesis models to generate speech of the users from the simulated conversations, integrating with noise sampled from MUSAN~\cite{musan} to simulate the background of the input, and reference speakers from Librispeech~\cite{librispeech} to get speaker-diverse synthesized speech corpus. This corpus is then fed into different FDSDS to obtain inference outputs, including speech and text, according to the modalities the models support. For the speech output, we applied Whisper~\cite{whisper} ASR model using WhisperX~\cite{whisperx} implementation to obtain speech transcription and Silero-VAD~\cite{Silero-VAD} to get utterance-level timestamps for further evaluation.
Finally, the benchmarking metrics, including Subjective and Objective counterparts, are produced using simple timestamp calculations and GPT/Llama inference.
\begin{table}[htp]
    \caption{Simulated Conversational Text Generated by GPT-4o model. We have various types of interruptions—\textbf{A}ffirmative Acknowledgment, \textbf{D}enial and Discontent, \textbf{F}urther Inquiry, Requiring a \textbf{R}epeat, and Topic \textbf{S}hift—to mirror real-life conversations.}
    \vspace{-0.9em}
    \centering\footnotesize
    \renewcommand{\arraystretch}{0.80}
    \begin{tabular}{c|c|c|c|c|c|c}
        \toprule
        \multirow{2}{*}{\# Convs}  & \multicolumn{5}{c|}{\textbf{Int}erruption Types} & \multirow{2}{*}{\# Total \textbf{Int}} \\
        & \# \textbf{A} & \# \textbf{D} & \# \textbf{F} & \# \textbf{R} & \# \textbf{S} &  \\
        \midrule
        293 & 350 & 58 & 393 & 109 & 316 & 1196 \\
        \bottomrule
    \end{tabular}
    \vspace{-1.9em}
    \label{tab:text}
\end{table}
\subsection{Dataset Design}
We design a corpus to simulate natural, full-duplex spoken dialogues between a user and an AI assistant. Our prompt for dialogue generation is inspired by~\cite{NEURIPS2024_duplex}, with some constraints to stabilize generation. Each conversation consists of up to 5 rounds, always initiated by the user, who is typically in a hurry and prone to interrupting. The dataset spans dozens of topics such as travel, food, entertainment, and fitness to broaden domain coverage.
\begin{table}[htp]
    \caption{User inquiry speech corpus. \textbf{E}asy / \textbf{M}edium / \textbf{H}ard interruption difficulties have diverse speech duration. For the noisy corpus, we generate three different SNR configurations for \textbf{E}asy portion of F5TTS and CosyVoice2.}
    \vspace{-0.9em}
    \centering\footnotesize
    \renewcommand{\arraystretch}{0.8}
    \begin{tabular}{c|l|c|c}
        \toprule
        ID & \makecell[c]{TTS Engines} & Duration (H) E/M/H & SNR  \\
        \midrule
        1 & ChatTTS & 4.2 / 2.8 / 2.3 & - \\
        2 & F5TTS & 3.8 / 2.4 / 2.0 & - \\
        3 & ~~~~~~+Noise-gap & 3.8 / - / - & 0/10/20 \\
        4 & ~~~~~~+Noise-bg & 3.8 / - / - & 0/10/20 \\
        5 & CosyVoice2 & 4.2 / 2.8 / 2.4 & - \\
        6 & ~~~~~~+Noise-gap & 4.2 / - / - & 0/10/20 \\
        7 & ~~~~~~+Noise-bg & 4.2 / - / - & 0/10/20 \\
        \bottomrule
    \end{tabular}
    \vspace{-1.9em}
    \label{tab:corpus}
\end{table}
The interruption statistics are shown in Table~\ref{tab:text}. The entire corpus includes 293 multi-round conversations with up to 4 interruptions each, resulting in around 1.2k interruptions, where sometimes affirmation and denial could happen simultaneously with further inquiry and topic shift. 

We only focus on the user part from the simulated conversations for the benchmarking and apply state-of-the-art TTS models, such as ChatTTS~\cite{2NoiseChatTTS}, F5TTS~\cite{F5TTS, SWividF5TTS}, and CosyVoice2~\cite{CosyVoice2,FunAudioLLMCosyVoice2} to generate a speaker-diverse user inquiry speech corpus. It's worth noting that we should keep the same speaker characteristics during the entire conversation to ensure a natural speech flow. Also, we inserted different lengths of silence between each inquiry speech segment to simulate varying extents of interruption. Specifically, we define three levels of interruptions, Easy, Medium, and Hard, where we insert random silences of 6–10 seconds, 4–6 seconds, and 2–4 seconds, respectively. Furthermore, we insert noise sampled from MUSAN with SNRs of 0, 10, and 20 to the entire corpus as background noise, namely \textbf{Noise-bg} and to only the silence gaps between each inquiry segment, namely \textbf{Noise-gap}, to evaluate the model's robustness to noisy conditions and noise interruptions. All speech is in 24KHz pcm format, with an overall duration of 40 hours. The statistics information for the generated corpus is shown in Table~\ref{tab:corpus}. 
\begin{table}[htp]
    \vspace{-0.5em}
    \caption{Models inference setup. The models are categorized into two classes based on FD implementations; built-in means the model itself handles FD events, while VAD-based means the model counts on an external VAD module to handle FD events. Chunk-size is measured in milliseconds. }
    \vspace{-0.9em}
    \centering\footnotesize
    \renewcommand{\arraystretch}{0.80}
    \begin{tabular}{p{1.4cm}|p{1.68cm}|c|c}
        \toprule
        \centering FD-Type & \centering Model & Chunk-size & VAD-thresh. \\
        \midrule
        \centering built-in & \centering Moshi & 80 & - \\
        \midrule
        \centering \multirow{2}*{VAD-based} & \centering Freeze-Omni & 107 & 0.8  
        \\
         & \centering VITA-1.5 & 200 & 0.7 \\
        \bottomrule
    \end{tabular}
    \vspace{-1.5em}
    \label{tab:models}
\end{table}

\begin{table*}[htp]
    \caption{The evaluation of model robustness under varying levels of user interruptions. Only the results from speech input simulated using Cosyvoice2 are reported in this paper. WER, SRR, SRIR, SIR, EIR, and C-PPL are assessed over the entire output speech response, while IRD, FSED, ERT, and EIT are reported using the \textbf{median} of all collected time values to mitigate the influence of outliers. The Score is calculated as the arithmetic \textbf{mean} of the six scores corresponding to six sub-perspectives derived from all successful responses. The Data name is formed as 5-\textbf{E}asy/\textbf{M}edium/\textbf{H}ard since we only report CosyVoice2-generated data in this table.}
    \vspace{-0.9em}
    \centering\footnotesize
    \renewcommand{\arraystretch}{0.80}
    \begin{tabular}{c|l|c|c|c|c|c|c|c|c|c|c|c}
        \toprule
        Model & Data & WER$\downarrow$ & SRR$\uparrow$ & SRIR$\uparrow$ & SIR$\uparrow$ & EIR$\downarrow$ & IRD$\downarrow$ & FSED$\downarrow$ & ERT & EIT & C-PPL & Score$\uparrow$ \\
        \midrule
        
        \multirow{3}*{Moshi} & 5-E & 5.3&61.7&79.4&83.1&27.9&1345&4155&221&2008 & 20 & 4.43 \\
        & 5-M & 5.0&45.6&78.4&77.1&23.1&1453&2735&235&1887 & 18 & 4.55 \\
        & 5-H & 5.0&34.1&75.3&73.0&19.5&1527&2020&235&1649 & 25 & 4.42 \\
        \midrule
        \multirow{3}*{Freeze-omni} & 5-E & - & 11.3&35.5&66.5&31.5&3618&515&180&2027 & 73 & 3.29\\
        & 5-M & - & 11.7&21.7&49.4&23.5&12200&449&259&2047 & 80 & 3.38 \\
        & 5-H & - & 11.2&18.3&49.0&23.7&11927&456&311&2093 & 110 & 3.14 \\
        \midrule
        \multirow{3}*{VITA-1.5} & 5-E & - & 26.9&54.7&75.7&25.1&13063&4242&234&2289 & 28 & 2.68 \\
        & 5-M & - & 18.3&48.5&81.9&32.5&10759&2601&176&2174 & 60 & 2.44 \\
        & 5-H & - & 17.1&45.7&78.0&35.1&4651&1840&253&2270 & 50 & 2.27 \\
        \bottomrule
    \end{tabular}
    \vspace{-1.5em}
    \label{tab:int-results}
\end{table*}
\subsection{Evaluation Metrics}

We evaluate reply quality using a combination of subjective and objective metrics. Specifically, we adopt the GPT-4o model as a scoring system that rates responses on six dimensions, including relevance, encouragement, simplicity, flexibility, practicality, and creativity, each on a 1–10 scale, then averages the sub-scores, resulting in the final subjective score. Simultaneously, we compute the conditioned perplexity (c-PPL) to quantify how well each reply aligns with the user request and conversation context. Together, these two approaches provide a holistic view of the quality of the system’s reply content.

\begin{figure}[htb]
     \vspace{+0.5em}
    \centering
    \includegraphics[width=1.00\linewidth]{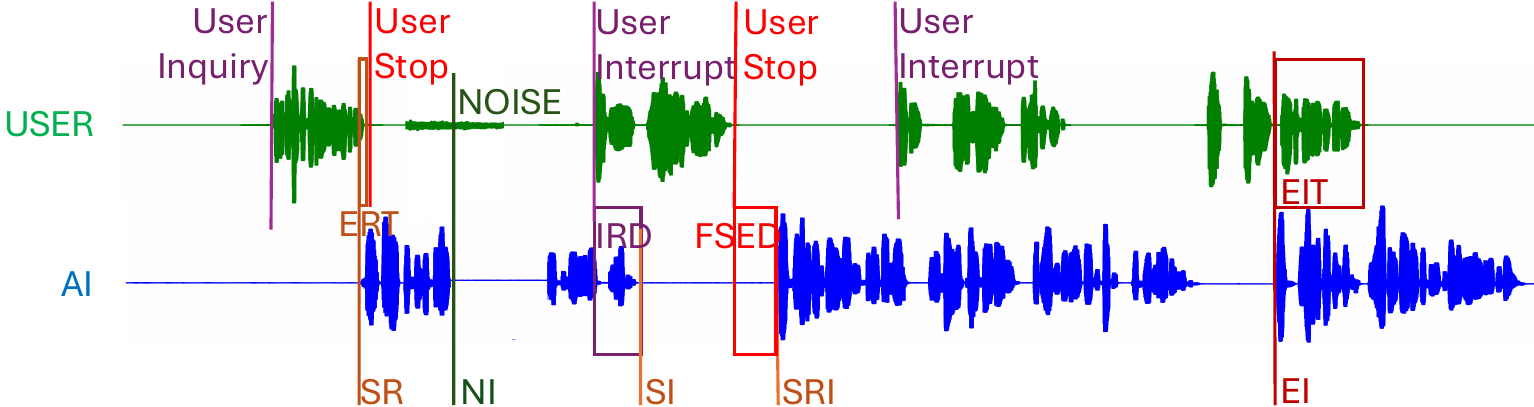}
    \caption{
    Visualization of real-time performance and interruption handling. In this example, AI successfully replies (\textbf{SR}) to the user inquiry with a slightly early reply time (\textbf{ERT}). Then, some noise from the user interrupts (\textbf{NI}) AI. AI resumes a reply, and the user interrupts, interruption response delays (\textbf{IRD}) the AI's being successfully interrupted (\textbf{SI}). When the user stops the interruption, AI successfully replies to interruption (\textbf{SRI}) after the first speech emits delay (\textbf{FSED}). However, AI does not respond to the user's next interruption and keeps talking. In the last round, AI early interrupts (\textbf{EI}) the user's inquiry for early interrupt time (\textbf{EIT}) before the user finishes. 
    }
    \label{fig:metrics-exp}
    \vspace{-1.9em}
\end{figure}

Beyond reply quality, we define several objective metrics to assess real-time performance and interruption handling, as shown in Fig.~\ref{fig:metrics-exp}. Interruption-related metrics—\textbf{SRR}ate (measures the odds of \textbf{S}uccess-\textbf{R}eplies per User-non-interrupt-inquiries), \textbf{SIR}ate (measures the rates of \textbf{S}uccess-\textbf{I}nterrupts per User-interrupt-inquiries), \textbf{SRIR}ate (measures the odds of \textbf{S}uccess-\textbf{R}eplies-to-\textbf{I}nterrupts per \textbf{SI}s), \textbf{EIR}ate (measures the odds of \textbf{E}arly-\textbf{I}nterrupts per User-inquiries), and \textbf{NIR}ate (measures the rates of \textbf{N}oise-\textbf{I}nterrupts per Noise-gaps between user inquiries)—capture how effectively the system processes user queries, interruptions, and noise events. 
Timing metrics—\textbf{IRD}(interrupt-response-delay), \textbf{FSED}(first-speech-emit-delay), \textbf{ERT}(early-reply-time), and \textbf{EIT}(early-interrupt-time)—measure delays and responsiveness, indicating how soon or prematurely the system starts or resumes talking. Lastly, WER (\textbf{W}ord-\textbf{E}rror-\textbf{R}ate) evaluates the accuracy of generated speech against output text, offering insight into the overall fidelity of the spoken output.

\section{Experiments}
\label{sec: exp}
We apply our benchmark pipeline to the State-of-the-Art open-source FDSDS that support English input and output, including Moshi~\cite{kyutaiMoshikoBF16}, Freeze-onmi~\cite{Freeze-omni,VITAMLLMFreezeOmni}, and VITA-1.5~\cite{VITA,VITAVITA-1.5}. 
\subsection{Setup}
Our experiments integrate robust computational resources with a server-client architecture, and each model operates on a single H20 GPU with 96GB VRAM. We deploy models using a development server code from official repositories, while clients are designed to simulate real-time speech streams. The evaluation settings for each model are shown in Table~\ref{tab:models}.
Specifically, we use the client to send audio with different chunk-sizes based on the server receiving settings. We follow the original VAD threshold settings of 0.8 and 0.7 for Freeze-omni and VITA-1.5.

We apply Silero-VAD with a threshold of 0.5 to get timestamp information from both the simulated user speech and the model's output. And we utilize Whisper-large-v3-turbo to perform ASR with a beam-size of 5. For subjective scoring, we call Openai API with a specially designed prompt using the GPT-4o-2024-11-20 model with the temperature set to 0.2. Finally, we use the Llama-3.3-70B-Instruct model to calculate Conditioned-PPL with the precision int-8, running on one single H20-96GB GPU.

\begin{table*}[htp]
    \caption{The evaluation of model robustness under varying background noise or sudden noise between user inquiries. SRR, SRIR, SIR, EIR, NIR, and C-PPL are assessed over the entire output speech response, while IRD, FSED, ERT, and EIT are reported using the \textbf{median} of all collected time values to mitigate the influence of outliers. The Score is calculated as the arithmetic \textbf{mean} of the scores for six sub-scoring perspectives derived from all successful responses. The data name is formed as \{ID\}-\{SNR\}.}
    \vspace{-0.9em}
    \centering\footnotesize
    \renewcommand{\arraystretch}{0.8}
    \begin{tabular}{c|l|c|c|c|c|c|c|c|c|c|c|c}
        \toprule
        Model & Data & SRR$\uparrow$ & SRIR$\uparrow$ & SIR$\uparrow$ & EIR$\downarrow$ & NIR $\downarrow$ & IRD$\downarrow$ & FSED$\downarrow$ & ERT & EIT & C-PPL & Score$\uparrow$ \\
        \midrule
        
        \multirow{6}*{Moshi} & 6-20 & 60.4&71.3&81.0&27.1&39.3&1577&5090&222&2049 & 16 & 4.38 \\
        & 6-10 & 63.8&78.9&84.6&24.8&46.0&1482&4410&231&1913 & 22 & 4.28 \\
        & 6-00 & 63.8&77.9&83.7&25.7&48.0&1638&4562&236&1936 & 27 & 4.26 \\ 
        & 7-20 & 59.7&74.9&83.9&25.4&37.6&1501&3077&207&2062 & 17 & 4.50 \\
        & 7-10 & 60.8&79.3&86.4&21.6&39.2&1530&1676&179&2157 & 20 & 4.24 \\
        & 7-0 & 59.7&77.3&87.8&22.4&41.6&1472&1311&182&2285 & 49 & 3.67 \\ 
        \midrule
        \multirow{6}*{Freeze-omni} & 6-20 & 14.0&34.1&61.1&27.0&6.5&11842&525&243&2069 & 142 & 3.15 \\
        & 6-10 & 13.8&35.6&65.0&28.9&6.5&4595&509&205&2174 & 77 & 3.22\\
        & 6-0 & 14.8&36.2&58.1&24.1&6.2&12491&554&281&2072 & 54 & 3.19\\
        & 7-20 & 9.9&31.0&61.0&26.6&3.6&4916&572&292&2103 & 66 & 3.01\\
        & 7-10 & 13.5&33.2&61.4&27.3&4.8&12254&633&224&2203 & 136 & 2.84\\
        & 7-0 & 11.9&35.0&63.5&26.4&4.4&11973&722&243&2134 & 95 & 2.36\\
        \midrule
        \multirow{6}*{VITA-1.5} & 6-20 & 30.6&55.9&81.1&29.5&16.2&10389&4308&269&2122 & 35 & 2.41\\
        & 6-10 & 32.3&61.0&84.9&32.8&21.1&9547&4226&271&2328 & 78 & 2.30\\
        & 6-0 & 32.6&61.4&85.7&34.2&21.9&9212&4162&323&2115 & 66 & 2.21\\
        & 7-20 & 28.8&56.8&81.3&28.1&16.2&12028&4541&280&2203 & 66 & 2.48 \\
        & 7-10 & 28.7&58.9&80.8&27.4&16.4&12218&4436&241&2303 & 73 & 2.24 \\
        & 7-0 & 30.2&59.1&82.6&28.4&17.4&12111&4501&255&2213 & 54 & 1.97 \\
        \bottomrule
    \end{tabular}
    \vspace{-1.9em}
    \label{tab:noisy-results}
\end{table*}
\subsection{Results}
We report the experimental results from two perspectives: the robustness to different levels of user interruptions and the robustness to varying levels of noisy conditions or sudden noise between user inquiries. Due to space limitations, here we report the results where we stream simulated speech corpus using CosyVoice2 into the systems only, which are ID 5-7 from table~\ref{tab:corpus}.

In table~\ref{tab:int-results}, we examine system performance under different frequencies of interruptions. We find that Moshi consistently outperforms VAD-based-interruption systems in interruption-related metrics, e.g., SRR, SRIR, and SIR, possibly due to suboptimal VAD thresholds used by VAD-based interruption systems, which might be prevented by predicting the possible interruption events internally in Moshi. 
While Moshi also has lower interruption delays, which are measured with IRD, FSED shows that Freeze-omni still replies much faster overall. 
Interestingly, all three systems sometimes respond a few hundred milliseconds earlier or interrupt the user’s inquiries by around two seconds, as seen in ERT, EIT, and EIR. This may be one of the training data designs of Moshi aimed at achieving faster response, whereas for VAD-controlled systems, it may be caused by the time shift introduced by the inaccuracy of the VAD module.

\begin{figure}[ht]
 
  \centering
  \begin{subfigure}{0.5\textwidth}
  \hspace{-2em}
    \centering
    \includegraphics[width=0.7\textwidth]{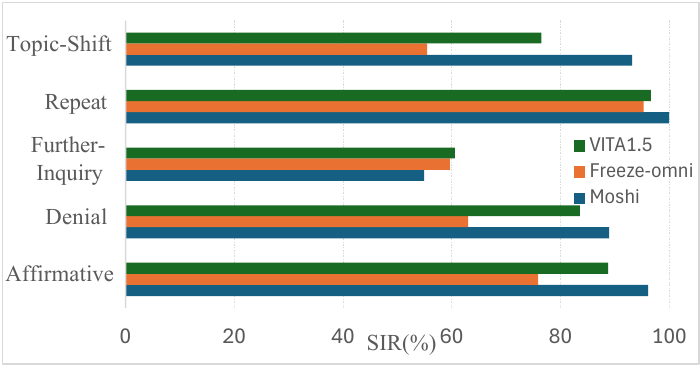}
  \end{subfigure}
  
  
  \begin{subfigure}{0.5\textwidth}
  \hspace{-2em}
    \centering
    \includegraphics[width=0.7\textwidth]{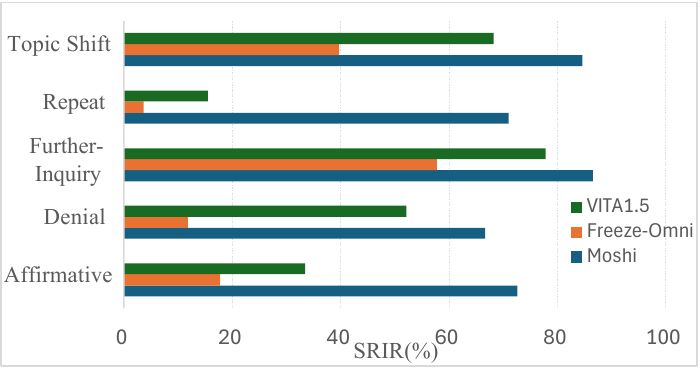}
  \end{subfigure}
  \caption{SIR and SRIR performance of the three models v.s. five types of interruption scenarios}
  \label{fig:SIR-SRIR}
  \vspace{-1.9em}
\end{figure}



When faced with more challenging interruption data, the WER for Moshi remains relatively stable, suggesting its generation quality is less affected by frequent interruptions. However, Moshi and VITA observe over 40\% relative drops in SRR with slight SRIR degradation, whereas Freeze-omni’s poor SRR results also accompany degraded SRIR in more challenging scenarios. 
Meanwhile, FSED and SIR reflect that frequent interruptions prompt Moshi to respond more quickly and make interruptions harder, possibly due to training data that includes similar speaker speaking styles and characteristics.
The notable response delays longer than 10 seconds in Freeze-omni and VITA-1.5 appear tied to VAD misjudgments and the instability of the official development server. 
Although C-PPL does not directly gauge response quality, it fluctuates more obviously for models using VAD under increased interruption difficulty, while Moshi remains relatively stable.
Finally, subjective scores evaluated from successful replies are stable for Moshi and Freeze-omni but degrade significantly for VITA-1.5 when interruptions become frequent. We also observe that VITA-1.5 randomly generates English-Chinese mixture speech, degrading the scores. These suggest that the quality of response under frequent interruption is not necessarily relevant to how the model handles interruptions; it's more related to the LLM's ability to understand and respond to user inquiries. 


In table~\ref{tab:noisy-results}, we evaluate system performance under noise situations. The results show that, under gap noise (data ID: 6-SNR) conditions, Moshi and VITA-1.5 exhibit an increasing NIR—while Freeze-omni remains relatively stable, likely due to a high VAD threshold of 0.8 that effectively filters most gap noise—and all models show stable interruption-related metrics (SRR, SRIR, SIR, and EIR) as the gap noise grows louder, except for VITA-1.5, which registers increases in these metrics, suggesting that a lower VAD threshold of 0.7 may trigger more false positives and fake interruptions. Meanwhile, the output quality score remains mainly unaffected by gap noise.

In contrast, with background noise (data ID: 7-SNR), although interruption-related metrics do not change significantly, Moshi’s FSED decreases markedly, possibly because more noise tokens are mistakenly interpreted as effective user input, and for all systems, subjective scores degrade significantly under low SNR conditions, which indicate that while the systems handle sudden noise relatively well, background noise poses challenges for noise filtering and robustness.

\section{Analysis} 
\label{sec: dis}

We designed five types of interruptions in our dataset. 
Here, we analyze the impact of different interruption types on FDSDS under clean and easy interruption settings (data ID: 5-E).
Fig.~\ref{fig:SIR-SRIR} shows the SIRs and SRIRs for the three models, broken down by each interruption type. In general, we can conclude that \textbf{Repeat}, where users would usually ask for a repeat of the model response, have the \textbf{Highest SIR} among all the models. However, it shows a relatively \textbf{Lower SRIR}, e.g., the worst for both Freeze-omni and VITA-1.5 and the second worst for Moshi, which suggests that \textbf{a \textit{pardon?} like request was not well-trained} to reply in these models. 
In contrast, \textbf{further inquiry} shows the \textbf{Worst SIR} with the \textbf{Best SRIR}, indicating that \textbf{a calm further question is challenging} for either model to detect, while such an inquiry is easier to reply for the LLMs.

\section{Conclusion and Future Work}
\label{sec: con}
In this work, we introduce a novel benchmarking pipeline that leverages LLMs, TTS, and ASR to comprehensively evaluate the performance of FDSDS, focusing on the model's performance under challenging interruption and noise interference scenarios. 
The evaluations generally include measures reflecting success or error rates across different user interruptions. Moreover, timing factors, including delays or early response, and quality aspects, such as scores rated by LLMs, assess real-time responsiveness, the naturalness of real-time interactions, and alignment with user demands.
Our benchmarking results show many challenges for FDSDS to remain under frequent interruptions and noisy environments. 
Hence, by applying our pipeline, future FDSDS should be able to evolve and become more natural in real-time human-machine interactions.
%

\section{Acknowledgement}
This research is supported by the RIE2025 Industry Alignment Fund – Industry Collaboration Projects (IAF-ICP) (Award I2301E0026), administered by A*STAR, as well as supported by Alibaba Group and NTU Singapore through Alibaba-NTU Global e-Sustainability CorpLab (ANGEL).

\bibliographystyle{IEEEtran}
\bibliography{mybib}

\end{document}